\documentclass[reprint,amsmath,amssymb]{revtex4-2}
\usepackage[utf8]{inputenc}
\usepackage[T1]{fontenc}
\usepackage{graphicx}
\usepackage[svgnames]{xcolor}
\usepackage{dcolumn}
\usepackage{bm}
\usepackage{enumerate}
\usepackage{hyperref}
\usepackage{mathcomp}
\usepackage{wasysym}
\usepackage{physics}
\usepackage{mathrsfs}
\hypersetup{
 colorlinks=true,%
 linkcolor=DarkSlateBlue,
 citecolor=DarkSlateBlue,
 urlcolor=DarkSlateBlue,
}
\everymath{\displaystyle}
\begin{document}

\preprint{}

\title{Exact vacuum solution with Hopf structure in general relativity}

\author{Junpei Harada}
 \email{jharada@hoku-iryo-u.ac.jp}
\affiliation{Health Sciences University of Hokkaido, 1757 Kanazawa, Tobetsu-cho, Ishikari-gun, Hokkaido 061-0293, Japan}

\date{June 25, 2025}

\begin{abstract}
An exact solution to the vacuum Einstein equations is presented, whose structure is based on the Hopf fibration. The solution employs a geodesic null vector field that defines a twisting congruence and appears in the metric in Kerr--Schild form. This solution is of Petrov type~D and involves two parameters. Remarkably, the resulting spacetime is regular, with no curvature singularities. Both the Kretschmann scalar and the Chern--Pontryagin scalar are nonzero and remain finite throughout the spacetime. In addition, the Newman--Penrose Weyl scalar $\Psi_2$ possesses both nonzero real and imaginary parts, reflecting the topologically nontrivial nature of the gravitational field. The spacetime also admits two Killing vector fields and a Killing--Yano tensor, which induces an associated Killing tensor, revealing its hidden symmetry. The derivation is simple and self-contained, offering a transparent and geometrically guided approach to finding new exact solutions in general relativity.
\end{abstract}

\maketitle

\section{Introduction\label{sec:intro}}
The Hopf fibration~\cite{Hopf1931}, mapping the 3-sphere $S^3$ onto $S^2$ with linked circular fibers, exhibits an elegant topological structure, whose influence spans diverse areas of physics~\cite{URBANTKE2003125}. In electromagnetism, for instance, field configurations with linked and knotted lines, known as hopfions, arise as exact solutions to the vacuum Maxwell equations~\cite{Ranada:1989wc,Kedia:2013bw}. These configurations have also been observed experimentally in materials. In gravity, these structures have been constructed in both linearized gravity and electromagnetism using a unified formalism~\cite{Smolka:2018rup}, and more recently, anti-self-dual spacetimes have been shown to be related to hopfions~\cite{Sabharwal:2019ngs}.
Such developments naturally raise the question: can general relativity, despite its inherent nonlinearity, admit exact solutions that reflect the topological richness of the Hopf fibration?

This work presents an exact solution to the vacuum Einstein equations, derived using the Kerr--Schild metric with a null vector field, built from the Hopf fibration. The resulting metric is of Petrov type D and includes two real parameters. It is free of curvature singularities, and both the Kretschmann and Chern--Pontryagin scalars are finite and nonzero throughout the spacetime. The solution also admits two Killing vectors and a nontrivial Killing--Yano tensor, which generates a corresponding Killing tensor and reveals the presence of hidden symmetries.

The construction is simple and self-contained, suggesting that topologically nontrivial geometries may be more accessible in general relativity than previously expected. The method employed here may offer a practical route to discovering further exact solutions in general relativity.

This paper is organized as follows. Section~\ref{sec:Kerr--Schild} presents the essential properties of Kerr--Schild metrics relevant to the construction. Section~\ref{sec:Hopf} derives a geodesic null vector field from the Hopf fibration. Section~\ref{sec:solution} presents the resulting exact solution. Section~\ref{sec:conclusion} summarizes the main features of the solution.

\section{Kerr--Schild metric\label{sec:Kerr--Schild}}
The Kerr--Schild metric~\cite{Kerr2009} takes the form
\begin{align}
	\mathrm{d} s^2 = g_{\mu\nu} \, \mathrm{d} x^\mu \mathrm{d} x^\nu = (\eta_{\mu\nu} + H k_\mu k_\nu) \, \mathrm{d} x^\mu \mathrm{d} x^\nu,
\end{align}
where $\eta_{\mu\nu}$ is the Minkowski metric, $H$ is a real function, and $k_\mu$ is a null vector with respect to both $g_{\mu\nu}$ and $\eta_{\mu\nu}$,
\begin{align}
	&g_{\mu\nu} k^\mu k^\nu = \eta_{\mu\nu} k^\mu k^\nu = 0,&
	&k^\mu = g^{\mu\nu}k_\nu = \eta^{\mu\nu}k_\nu.&
\end{align}
The inverse of the metric tensor depends linearly on $H$
\begin{align}
	g^{\mu\nu} = \eta^{\mu\nu} - H k^\mu k^\nu,
\end{align}
which is an exact equation. The determinant of the Kerr--Schild metric satisfies $\det (g_{\mu\nu}) = \det (\eta_{\mu\nu})$. Thus, the Kerr--Schild metric is uniquely determined once the null vector $k_\mu$ and a real function $H$ are specified.

Several exact solutions to the vacuum Einstein equations can be expressed in the Kerr--Schild form. 
In coordinates $(t, x, y, z)$ with $\eta_{\mu\nu}=\text{diag}(-1,1,1,1)$, 
the Schwarzschild metric takes the form
\begin{align}
	&k_\mu = \qty( 1,  \frac{x}{r}, \frac{y}{r}, \frac{z}{r}),&
	&H = \frac{2M}{r},&
\end{align}	
and the Kerr metric~\cite{Kerr:1963aa} is given by
\begin{align}
	k_\mu = \qty( 1,  \frac{rx+ay}{r^2+a^2}, \frac{ry-ax}{r^2+a^2}, \frac{z}{r}),\,\,
	H = \frac{2M r^3}{r^4+a^2 z^2},
\end{align}
where the parameter $r$ is not a coordinate but is implicitly determined by the null condition $k_\mu k^\mu=0$ in both the Schwarzschild and the Kerr metrics. 

In light-cone coordinates, $x^0=v=(t+z)/\sqrt{2}$, $x^1=u=(t-z)/\sqrt{2}$, $x^2=x$, $x^3=y$ with $\eta_{01}=\eta_{10}=-1$ and $\eta_{22}=\eta_{33}=1$, the pp-wave spacetime takes the form
\begin{align}
	&k_\mu = \qty(0, 1, 0, 0),&
	&H_{,xx}+H_{,yy}=0,&
\end{align}
where $H=H(u,x,y)$ and $H_{,xx}=\partial^2 H/\partial x^2$.

In all these examples---Schwarzschild, Kerr, and pp-wave---the null vector $k_\mu$ is (affinely parametrized) geodesic with respect to both $g_{\mu\nu}$ and $\eta_{\mu\nu}$, satisfying
\begin{align}
	k^\nu \nabla_\nu k_\mu = k^\nu \partial_\nu k_\mu = 0,
\end{align}
where $\nabla_\mu$ is a covariant derivative. 

\section{Hopf-based null vector\label{sec:Hopf}}
The Hopf fibration~\cite{Hopf1931} is used to construct the null vector $k_\mu$ as follows. Two complex numbers $z_1, z_2 \in\mathbb{C}$ satisfying $|z_1|^2+|z_2|^2=1$ can be parametrized as
\begin{align}
	&z_1 =\frac{2(x + i y)}{r^2 - t^2 + 1 + 2it},&
	&z_2 =\frac{r^2 - t^2 - 1 - 2 i z}{r^2 - t^2 + 1 + 2it},&
\end{align}
where $r^2 = x^2 + y^2 + z^2$. A pair of complex numbers $(z_1, z_2)$ defines coordinates on the 3-sphere $S^3$. At each point on $S^3$, consider the following map:
\begin{align}
	(z_1, z_2) \rightarrow \frac{z_1}{z_2} = \frac{2(x + i y)}{r^2 -t^2 - 1 - 2 i z} =: \phi (t,x,y,z),
\end{align}
where $\phi$ is a complex scalar. Since $z_1/z_2=e^{i\alpha}z_1/(e^{i\alpha} z_2)$ for $\alpha=[0,2\pi)$,  every point on the circle $e^{i\alpha}(z_1, z_2)$ in $S^3$ maps to a single point $z_1/z_2=\phi$. Therefore, the inverse map from $z_1/z_2$ to the points on $S^3$ forms an $S^1$ fiber~\cite{Hopf1931}. 

A tangent vector $k^\mu$ to this $S^1$ fiber can be defined by
\begin{align}
	F_{\mu\nu} k^\nu = 0,
\end{align}
where
\begin{align}
	F_{\mu\nu} := \frac{1}{i}\qty(\partial_\mu \phi^* \partial_\nu \phi - \partial_\nu \phi^* \partial_\mu \phi).
\end{align}
An imaginary unit $i$ is introduced so that $F_{\mu\nu}$ is real. The vector $k_\mu$ is also required to satisfy both the null condition $\eta^{\mu\nu}k_\mu k_\nu=0$ and the geodesic condition $k^\nu \partial_\nu k_\mu=0$. One finds that the following $k_\mu$ satisfies these conditions:
\begin{align}
	k_\mu &=
	\frac{1}{\sqrt{2}\qty(1+(t-z)^2)}
	\begin{pmatrix}1 +(t-z)^2+ x^2 + y^2\\
	- 2 (t - z) x - 2y \\
	- 2 (t - z) y + 2 x \\
	1+(t-z)^2 - x^2 - y^2
	\end{pmatrix},
	\label{eq:kernel_Cartesian}
\end{align}
where $t-z$ direction has been chosen rather than $t+z$, and a factor $1/\sqrt{2}$ is introduced for later convenience in light-cone coordinates. The one-form $k_\mu \mathrm{d} x^\mu$ is invariant under rotations in the $xy$-plane. 

\begin{figure*}
\includegraphics[width=150mm]{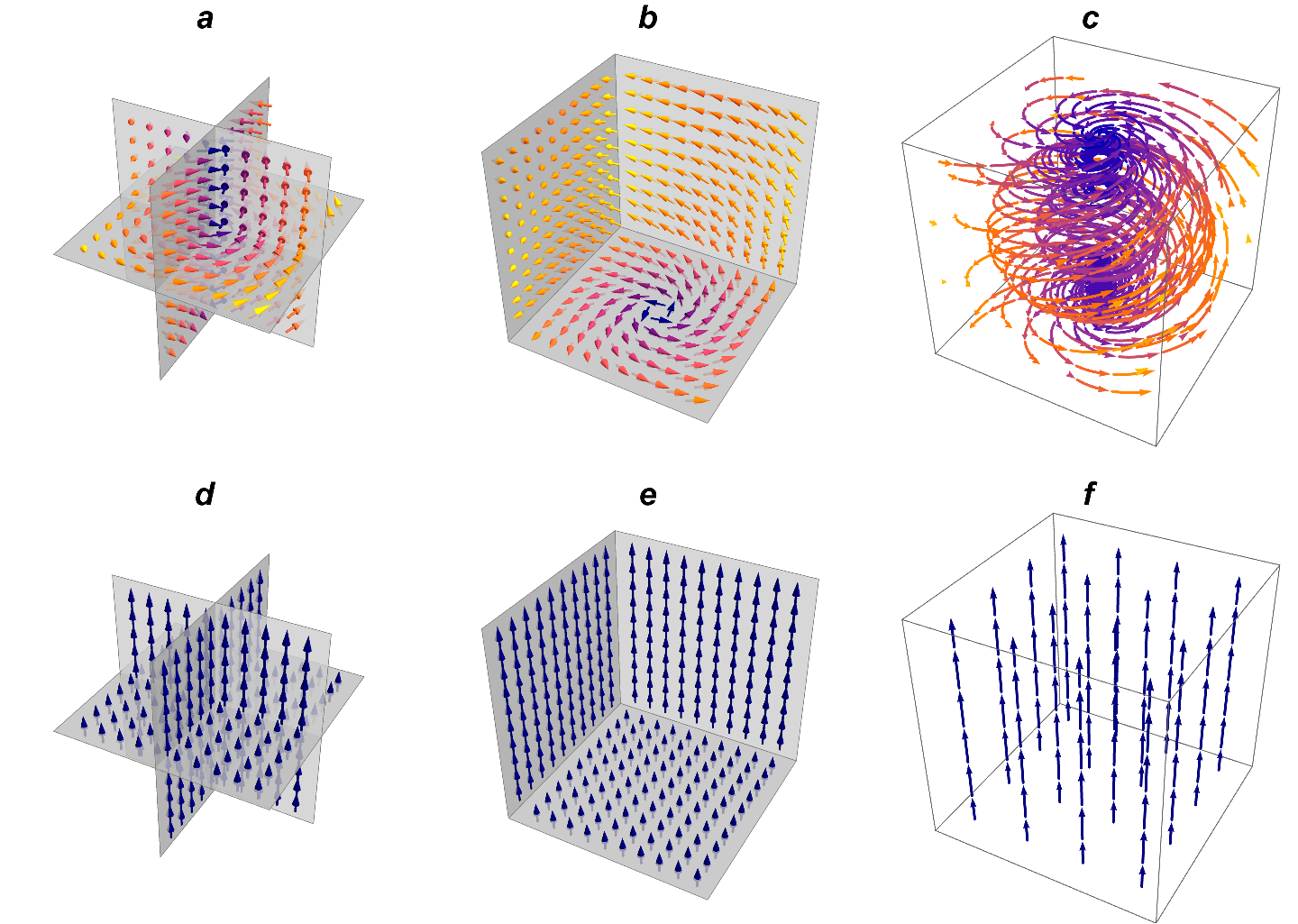}%
\caption{\label{fig:kernel}{\bf Geodesic null vector fields (a, b, d, e) and their integral curves (c, f).} Panels (a)--(c) correspond to the present solution [Eq.~\eqref{eq:kernel}], and panels (d)--(f) to the pp-wave. The vertical axis represents the $u$-axis. Color indicates the magnitude of the vector in Euclidean $(u, x, y)$ space (brighter colors denote larger magnitudes).}
\end{figure*}

Since $k_\mu$ depends on $t-z$, it is convenient to adopt light-cone coordinates. For $x^0=v=(t+z)/\sqrt{2}$, $x^1=u=(t-z)/\sqrt{2}$, $x^2=x$, $x^3=y$, the null vector $k_\mu$ is given by
\begin{align}
	k_\mu = \qty(1, \frac{x^2+y^2}{2 u^2 + 1}, - \frac{2 u x + \sqrt{2} y}{2 u^2 + 1}, - \frac{2 u y - \sqrt{2}x}{2 u^2 + 1}).
	\label{eq:kernel}
\end{align}
Figure~\ref{fig:kernel} compares $\bm{k}=(k_u, k_x, k_y)$ with the constant vector $\bm{k}=(1,0,0)$ representing a pp-wave. In the figure, the vertical axis corresponds to the $u$-axis. It is evident that the null vector in Eq.~\eqref{eq:kernel} exhibits a twisted structure that is qualitatively distinct from that of a pp-wave.

\section{Regular Petrov D spacetime\label{sec:solution}}
The $k_\mu$ from Eq.~\eqref{eq:kernel} is used in the Kerr--Schild metric
\begin{align}
	\mathrm{d} s^2 
	= -2  \mathrm{d} u \mathrm{d} v+ \mathrm{d} x^2 + \mathrm{d} y^2
	+H k_\mu k_\nu \, \mathrm{d} x^\mu \mathrm{d} x^\nu.
	\label{eq:Hopf_metric}
\end{align}
If $H$ is a function of $u$, $H=H(u)$, then two components of Ricci tensor contain only first-order derivatives
\begin{align}
	R^2{}_2 = R^3{}_3 = \frac{2}{2 u^2 + 1}\qty(u H^\prime (u) + \frac{2 u^2 -1}{2 u^2 + 1}H(u)), 
\end{align}
where $H^\prime (u)=\mathrm{d} H(u)/\mathrm{d} u$. Solving $R^2{}_2=0$, one obtains
\begin{align}
	H(u) = \frac{N u}{2 u^2 + 1}
	\label{eq:scale}
\end{align}
where $N$ is an integration constant. If $H(u)$ is given by this function, all components of the Ricci tensor vanish. Therefore, the metric in Eq.~\eqref{eq:Hopf_metric}, together with Eqs.~\eqref{eq:kernel} and~\eqref{eq:scale}, represents an exact solution to the vacuum Einstein equations.

By restoring a dimensional parameter, one obtains
\begin{align}
	k_\mu &= \qty(1, \frac{x^2+y^2}{2\qty(u^2 + b^2)}, - \frac{u x + b y}{u^2 + b^2}, - \frac{u y - b x}{u^2 + b^2}),\label{eq:kernel_dim}
\end{align}
and
\begin{align}
	H(u) &= \frac{N u}{u^2 + b^2},
\end{align}
where $b > 0$ is a parameter with the dimension of length. Note that $b < 0$ can be absorbed by redefining $x$ and $y$, so $b>0$ is assumed without loss of generality, and $b = 0$ is excluded by construction. The integration constant $N$ is redefined accordingly. It is also worth noting that the gauge field $A_\mu:=H(u) k_\mu$ satisfies the vacuum Maxwell equations in flat spacetime, $\eta^{\mu\rho}\partial_\rho (\partial_\mu A_\nu - \partial_\nu A_\mu)=0$.

The final form of the solution is thus
\begin{align}
	&\mathrm{d} s^2=\, -2 \mathrm{d} u \mathrm{d} v+ \mathrm{d} x^2 + \mathrm{d} y^2 + \frac{N u}{u^2 + b^2}\nonumber\\
	&\times \qty[\mathrm{d} v+\frac{x^2+y^2}{2\qty(u^2 + b^2)} \, \mathrm{d} u - \frac{u x + b y}{u^2 + b^2} \, \mathrm{d} x - \frac{u y - b x}{u^2 + b^2} \, \mathrm{d} y]^2.
	\label{eq:solution1}	
\end{align}
In the limit $N \rightarrow 0$ or $u\rightarrow \pm \infty$, the solution asymptotically approaches the Minkowski metric.

This solution admits two Killing vector fields $\xi^\mu$, given in coordinates $(v, u, x, y)$ by
\begin{align}
	\xi^\mu = \qty(1, 0, 0, 0),
	\, \,
	\qty(0, 0, y, -x)
\end{align}
and an antisymmetric Killing--Yano tensor $f_{\mu\nu}$,
\begin{align}
	f_{\mu\nu} = 
	\begin{pmatrix}
		0 & b & 0 & 0 \\
		-b & 0 & y & -x \\
		0 & -y & 0 & u \\		
		0 & x & -u & 0 \\				
	\end{pmatrix}
\end{align}
which satisfies $\nabla_\mu f_{\nu\rho} + \nabla_\nu f_{\mu\rho}=0$. From this Killing--Yano tensor, one can construct the Killing tensor $K_{\mu\nu}$,
\begin{align}
	K_{\mu\nu} := f_{\mu \rho}f_{\nu\sigma} g^{\rho\sigma}.
\end{align}
This tensor $K_{\mu\nu}$ satisfies $\nabla_{\rho}K_{\mu\nu} + \nabla_\mu K_{\nu\rho} + \nabla_\nu K_{\rho\mu}=0$,
and it is irreducible, in the sense that it cannot be written as a linear combination of $\xi_{(\mu} \xi_{\nu)}$ and the metric $g_{\mu\nu}$.

Since $R_{\mu\nu}=0$, the Kretschmann scalar $R_{\mu\nu\rho\sigma}R^{\mu\nu\rho\sigma}$ equals the square of the Weyl tensor,
\begin{align}
	R_{\mu\nu\rho\sigma}R^{\mu\nu\rho\sigma}  
	&= C_{\mu\nu\rho\sigma}C^{\mu\nu\rho\sigma}  \nonumber\\
	&= \frac{12N^2(u^6 - 15 b^2 u^4 + 15 b^4 u^2 - b^6)}{(u^2 + b^2)^6},
\end{align}
and the Chern--Pontryagin scalar ${^\star{}R_{\mu\nu\rho\sigma}}R^{\mu\nu\rho\sigma}$ is
\begin{align}
	{^\star{}R_{\mu\nu\rho\sigma}}R^{\mu\nu\rho\sigma} 
	&= {^\star{}C_{\mu\nu\rho\sigma}}C^{\mu\nu\rho\sigma} \nonumber\\
	&= \frac{24 N^2 b u (3 u^4 - 10 b^2 u^2 + 3 b^4)}{(u^2+b^2)^6},
\end{align}
where ${^\star{}R_{\mu\nu\rho\sigma}}=(1/2)\epsilon_{\mu\nu\alpha\beta}R^{\alpha\beta}{}_{\rho\sigma}$. 

The Newman--Penrose Weyl scalars are
\begin{align}
	&\Psi_2 = - \frac{N}{2 (u + i b)^3},&
	&\Psi_0=\Psi_1=\Psi_3=\Psi_4=0.&
\end{align}
This implies that the solution is of Petrov type~D, as in the Schwarzschild and Kerr metrics. Both the real and imaginary parts of $\Psi_2$ are nonzero, as in the Kerr case. 

Using the Newman--Penrose scalars, the curvature invariants can be expressed in particularly simple forms
\begin{alignat}{3}
	I_1&=C_{\mu\nu\rho\sigma}C^{\mu\nu\rho\sigma}
	&&= + 48 \Re \Psi_2^2 &&= + \Re \frac{12 N^2}{(u + i b)^6},\\
	I_2&={^\star{}C_{\mu\nu\rho\sigma}}C^{\mu\nu\rho\sigma} 
	&&= - 48 \Im \Psi_2^2 &&= -\Im \frac{12 N^2}{(u + i b)^6},
\end{alignat}
which follow from the identity~\cite{Cherubini:2002gen}
\begin{align}
	I_1 - i I_2 =16\qty(3 \Psi_2^2 + \Psi_0 \Psi_4 - 4 \Psi_0 \Psi_3).
\end{align}
The magnitude $\qty| I_1 - i I_2|$ is 
\begin{align}
	\sqrt{(I_1)^2 + (I_2)^2} = 48 |\Psi_2|^2 = \frac{12 N^2}{(u^2 + b^2)^3}.
\end{align}
This expression is finite and nonzero everywhere, with no curvature singularities since $b > 0$. It depends only on $u$ and is independent of $v, x, y$, indicating a wave propagating along the $+z$-direction at the speed of light with a planar wavefront. Due to the $u^{-6}$ dependence, the curvature is concentrated near the origin of the $u$-axis and falls off rapidly away from it.

Table~\ref{tab:list} summarizes several solutions in the Kerr--Schild form.

\begin{table*}
\caption{\label{tab:list} 
A list of exact solutions to the vacuum Einstein equations in Kerr--Schild form, $\mathrm{d} s^2 = (\eta_{\mu\nu} + H k_\mu k_\nu) \, \mathrm{d} x^\mu \mathrm{d} x^\nu$, where $\eta_{\mu\nu}$ is the Minkowski metric, $H$ is a real scalar function, and $k_\mu$ is a geodesic null vector. (1) Name of the solution. (2) One-form $k_\mu \mathrm{d} x^\mu$. Cartesian coordinates are used for the Schwarzschild and Kerr solutions, while light-cone coordinates, $u=2^{-1/2}(t-z), v=2^{-1/2}(t+z)$, are used for the pp-wave and the present work. (3) The scalar function $H$, or the condition it satisfies. (4) Nonzero components of the Newman--Penrose Weyl scalars (all others vanish). (5) Presence or absence of scalars, $I_1 = C_{\mu\nu\rho\sigma} C^{\mu\nu\rho\sigma}$ and $I_2={^\star{}C_{\mu\nu\rho\sigma}}C^{\mu\nu\rho\sigma}$. (6) Petrov classification. (7) Presence or absence of curvature singularities.}
\begin{ruledtabular}
\begin{tabular}{lcccccc}
 Name& $k_\mu \mathrm{d} x^\mu$ & $H$ & Weyl & $I_1, I_2$ & Type & Regularity \\ 
 (1) & (2) & (3)& (4) & (5) & (6) & (7)\\\hline
 Schwarzschild & $\vphantom{\dfrac{1}{1}} \mathrm{d} t + \frac{x}{r} \, \mathrm{d} x + \frac{y}{r} \, \mathrm{d} y + \frac{z}{r} \, \mathrm{d} z{}~\footnote{In both the Schwarzschild and Kerr solutions, the parameter $r$ is implicitly determined by the null condition $g^{\mu\nu}k_\mu k_\nu = \eta^{\mu\nu}k_\mu k_\nu = 0$.} \vphantom{\dfrac{1}{1}}$ & $2M/r$ & $\Psi_2$ & $I_1 \not=0$, $I_2=0$ & D & Singular\\
 Kerr&$\mathrm{d} t + \frac{rx+ay}{r^2+a^2} \,\mathrm{d} x+\frac{ry - ax}{r^2+a^2} \, \mathrm{d} y + \frac{z}{r} \, \mathrm{d} z$~\footnotemark[1] &$\frac{2Mr^3}{r^4+a^2 z^2}$ & $ \Psi_2$ & $I_1 \not=0$, $I_2\not=0$ & D & Singular\\
 pp-wave&$\mathrm{d} u$ & $H_{,xx}+H_{,yy}=0$~\footnote{The function $H$ depends on $u$, $x$, and $y$, i.e., $H=H(u,x,y)$.} & $ \Psi_4$ & $I_1 =I_2=0$ & N & Regular \\
 Present work& $\mathrm{d} v+\frac{x^2+y^2}{2(u^2+b^2)} \, \mathrm{d} u - \frac{u x + b y}{u^2+b^2} \, \mathrm{d} x - \frac{u y - b x}{u^2+b^2} \, \mathrm{d} y$~\footnote{The parameter $b$ has the dimension of length and satisfies $b>0$.} & $\frac{Nu}{u^2+b^2}$ &  $\Psi_2$ & $I_1 \not=0$, $I_2\not=0$ & D & Regular\\
\end{tabular}
\end{ruledtabular}
\end{table*}

Next, consider the case where $H = H(u, x, y)$. When $H$ depends on $u$, $x$, and $y$, the Einstein equations become more complicated, though they remain potentially tractable since they are still linear in $H(u, x, y)$. Moreover, three of the Newman--Penrose Ricci scalars---the real scalar $\Phi_{22}$ and the complex $\Phi_{02}$ and $\Phi_{12}$---vanish
\begin{align}
	\Phi_{22} = \Phi_{02} = \Phi_{12} = 0.
\end{align}
The remaining Ricci scalars---two real ($\Phi_{00}$, $\Phi_{11}$), one complex ($\Phi_{01}$), and the Ricci scalar $R$---are nonzero. 

Among these, $\Phi_{00}$ yields the Laplace equation
\begin{align}
	H_{,xx} + H_{,yy} = 0.
\end{align}
This implies that $H$ must be a harmonic function. The remaining quantities $\Phi_{11}$, $\Phi_{01}$, and the Ricci scalar $R$ are  complicated and are not analyzed further in the present work. As for the Newman--Penrose Weyl scalars, $\Psi_0 = \Psi_1 = 0$, while the others are generally nonvanishing. Therefore, it is possible that gravitational wave solutions of Petrov type II may exist in this setting, though the existence of such solutions remains an open question.

Finally, the following metric, with curvature propagating along the $-z$ direction, is clearly an exact solution,
\begin{align}
	&\mathrm{d} s^2=\, -2 \mathrm{d} u \mathrm{d} v+ \mathrm{d} x^2 + \mathrm{d} y^2 + \frac{N v}{v^2 + b^2}\nonumber\\
	&\times \qty[\frac{x^2+y^2}{2\qty(v^2 + b^2)}\, \mathrm{d} v + \mathrm{d} u - \frac{v x + b y}{v^2 + b^2} \, \mathrm{d} x - \frac{v y - b x}{v^2 + b^2} \, \mathrm{d} y]^2.
	\label{eq:solution2}
\end{align}
If a solution combining Eqs.~\eqref{eq:solution1} and~\eqref{eq:solution2} can be found, it would allow the study of their interactions and stabilities. Although constructing such a solution is nontrivial due to the nonlinearity of the Einstein equations, this proposal may inspire further investigation.\\

\section{Conclusions\label{sec:conclusion}}
This work has presented an exact vacuum solution to the Einstein field equations in Kerr--Schild form, constructed using a geodesic null vector field derived from the Hopf fibration. The vector field defines a twisting null congruence with nontrivial topology. The solution involves two parameters and is of Petrov type~D, with a nonvanishing Weyl scalar $\Psi_2$ that possesses both real and imaginary components. The resulting spacetime is regular everywhere, with no curvature singularities. Both the Kretschmann scalar and the Chern--Pontryagin scalar are finite and nonzero, indicating the presence of a genuinely nonlinear and topologically nontrivial gravitational field. The spacetime also admits two Killing vector fields and a nontrivial Killing–Yano tensor, indicating hidden geometric symmetries.

To the best of current knowledge, this appears to be the first explicit example of a singularity-free, vacuum, exact solution of  Petrov type~D constructed from a Hopf-structured null vector field. Notably, this metric is not listed in the standard reference~\cite{Stephani:2003tm}, and its derivation presented here is remarkably simple and fully self-contained. The construction demonstrates that geometrically motivated null structures---particularly those with nontrivial topology, such as the Hopf fibration---can serve as powerful tools for discovering exact solutions in general relativity. This suggests that further exploration of such structures may reveal new classes of physically relevant spacetimes beyond the known catalog.

\begin{acknowledgments}
This work was supported by JSPS KAKENHI Grant No. JP22K03599.
\end{acknowledgments}

\bibliography{references}
\end{document}